# Phase Diagram of Metal-Insulator Transition in System with Anderson-Hubbard Centers

Yu. Skorenkyy, L. Didukh, O.Kramar and Yu. Dovhopyaty

*Ternopil National Technical University, Physics Department, 56 Ruska Str., Ternopil, Ukraine*

The model of a strongly correlated system in which periodically spaced Anderson-Hubbard centers are introduced into narrow-band metal is considered. Besides the interactions between localized magnetic moments and strong on-site Coulomb interaction, the model takes into account the hybridization of localized and band states. To study the efect of the lattice deformation on the electrical properties of the system the phonon term and elastic energy have been taken into account. Green functions for band and localized electrons have been found. On this base, the energy spectrum has been investigated as function of model parameters, temperature and external pressure. The criterion of metal-insulator transition for integer value of electron concentration has been derived and the phase diagram of the metal-insulator transition has been built.

## 1. Introduction

In recent decades, the progress of technology is connected with the synthesis of new materials with unique electric al and magnetic properties. Materials with Anderson-Hubbard centers implemented in a matrix of narrow-band conductor are amongst the most perspective ones. The first attempt to built a theoretical description of such impurity center nas been done in the pioneering paper [1]. In papers [2, 3] a generalization of single impurity Anderson model [1] for the case of a system of periodically spaced Anderson-Hubbard centers has been proposed. In this periodic Anderson model, in the limit of strong intrasite interaction between localized magnetic moments, the indirect exchange through the band electron subsystem occurs. In papers [4, 5] the single impurity and double impurity Anderson models have been used to study the electric conductance of the systems with quantum dots and the magnetic ordering in band electron subsystem was found crucially important for the spin-dependent transport through a quantum dot. In this paper, an effective Hamiltonian taking into account basic interactions in the localized electron subsystem as well as hybridization between the localized and band electrons is used to study electrical properties of Anderson-Hubbard material.

## 2. The model Hamiltonian

We start from the model of Anderson-Hubbard material which generalizes the models [2, 6] and take into account the peculiarities of correlation effects in narrow energy bands. The Hamiltonian contains terms describing localized (*d*) subsystem and band (*s*) subsystem as well as their hybridization

$$H = \sum_{k\sigma}(\varepsilon_k - \mu)c^+_{k\sigma}c_{k\sigma} + (E_d - \mu)\sum_{i\sigma}d^+_{i\sigma}d_{i\sigma} +$$
$$\sum_{ij\sigma}t_{ij}(n)d^+_{i\sigma}d_{j\sigma} + \sum_{ij\sigma}\left(T(ij)d^+_{i\sigma}d_{j\sigma}n_{i\bar\sigma} + h.c.\right) +$$
$$+ U\sum_i n_{i\uparrow}n_{i\downarrow} + \frac{1}{2}\sum_{ij\sigma\sigma'}J(ij)d^+_{i\sigma}d^+_{j\sigma'}d_{i\sigma'}d_{j\sigma} + \quad (1)$$
$$+ \sum_{ik\sigma}\left(V(i\mathbf{k})d^+_{i\sigma}c_{\bar k\sigma} + h.c.\right) +$$
$$+ \sum_{\substack{ij\mathbf{k}\\ i\ne j}}\left(V(ij\mathbf{k},\mathbf{k})d^+_{i\sigma}d^+_{j\bar\sigma}c_{-\mathbf{k}\bar\sigma}c_{\mathbf{k}\sigma} + h.c.\right),$$

where $\varepsilon_{\vec k}$ denotes the energy of band electron with wave vector $\mathbf{k}$; $V(i\mathbf{k})$ and $V(ij\mathbf{k},-\mathbf{k})$ are matrix elements describing single-electron and two-electron hybridization of band and localized states; $d^+_{i\sigma}$, $d_{i\sigma}$ are creation and annihilation operators for spin $\sigma$ electron on $i^{th}$ center in localized (*d*) state; $c^+_{\mathbf{k}\sigma}$, $c_{\vec k\sigma}$ are operators of band electron creation and annihilation. Hopping of electrons in *d*-subsystem is strongly influenced by correlated hopping (we take into account two possible mechanisms of correlated hopping which, effectively, cause the energy subbands narrowing). The model Hamiltonian takes into account basic processes and interactions in a narrow non-degenerate band, namely electron hoppings (the third and fourth sums in (1)), intra-site Coulomb repulsion (the fifth sum), interatomic exchange (the sixth sum). The terms "localized" and "band" used here can have different sense depending on the peculiarities of the material under consideration. If a transition metal is studied then the localized subsystem are 3*d*-electrons and band subsystem is formed by *s-p*-electrons. For the case of narrow band oxides, 3*d* electrons form the localized sybsystem and band states correspond to both 3*d* electrons of transition metal and 2*p* of oxygen sybsystem, in rare earth compounds one has localized *f*-electrons and band *s-p-d*-electrons.

## 3. Metal-insulator transition

Let us consider a partial case of the model

$$H = -\mu \sum_{i\sigma}\left(c^+_{i\sigma}c_{i\sigma} + d^+_{i\sigma}d_{i\sigma}\right) + E_d \sum_{i\sigma} d^+_{i\sigma}d_{i\sigma} +$$
$$+ E_b(u)\sum_{i\sigma} c^+_{i\sigma}c_{i\sigma} + U\sum_i n_{i\uparrow}n_{i\downarrow} + H_s + H_{sd} + \quad (2)$$
$$+ \sum_{\bar{q}f}\hbar\omega_f(\bar{q})b^+_{\bar{q}f}b_{\bar{q}f} + \frac{1}{2}NV_0C\bar{u}^2,$$

$$H_s = \sum_{ij\sigma} t_{ij}(u) c^+_{i\sigma}c_{i\sigma}, \quad (3)$$

$$H_{sd} = V(u)\sum_{i\sigma}\left(c^+_{i\sigma}d_{i\sigma} + d^+_{i\sigma}c_{j\sigma}\right), \quad (4)$$

with hybridization of band and localized states and take into consideration the effect of lattice strain $u$ on the electronic subsystem, where

$$t_{ij}(u) = t_{ij}\left(1 + \frac{BV_0}{2w}u\right), \quad V(u) = V - gu, \quad 2w$$

represents unperturbed energy band width, $b_{\bar{q}f}$ operators describe phonon subsystem and values of parameters $g$, $B$, $V_0$, $C$ depend on the narrow-band compound [7].

We use the Green function method for calculation and write the equation for localised electron and band electron Green functions as

$$\langle\langle c_{p\uparrow}|c^+_{p'\uparrow}\rangle\rangle(E + \mu - E_b(u)) = \frac{\delta_{pp'}}{2\pi} +$$
$$+ \langle\langle[c_{p\uparrow};H_s]|c^+_{p'\uparrow}\rangle\rangle + \langle\langle[c_{p\uparrow};H_{sd}]|c^+_{p'\uparrow}\rangle\rangle,$$

$$\langle\langle d_{p\uparrow}|d^+_{p'\uparrow}\rangle\rangle(E + \mu - E_d) - U\langle\langle n_{p\downarrow}d_{p\uparrow}|d^+_{p'\uparrow}\rangle\rangle =$$
$$= \frac{\delta_{pp'}}{2\pi} + \langle\langle[d_{p\uparrow};H_{sd}]|d^+_{p'\uparrow}\rangle\rangle.$$

To break off the chain of equation we apply a projection procedure [8]

$$[c_{p\uparrow};H_{sd}] = \sum_i \varepsilon_{pj}d_{j\uparrow}; \quad (5)$$

$$[d_{p\uparrow};H_{sd}] = \sum_j \xi_{pj}c_{j\uparrow}, \quad (6)$$

$$\langle\langle n_{p\downarrow}d_{p\uparrow}|d^+_{p'\uparrow}\rangle\rangle \cong \langle n_{p\downarrow}\rangle\langle\langle d_{p\uparrow}|d^+_{p'\uparrow}\rangle\rangle \quad (7)$$

and analogous decouplings in the equations for functions $\langle\langle c_{p\uparrow}|d^+_{p'\uparrow}\rangle\rangle$ and $\langle\langle d_{p\uparrow}|c^+_{p'\uparrow}\rangle\rangle$.

Solving the equations with respect to band and localised electrons Green functions we obtain the energy spectrum

$$E_{1,2} = -\mu + \frac{E_d + E_b(u)}{2} + \frac{U\langle n_{p\downarrow}\rangle}{2} + \frac{t_{\bar{k}}(u)}{2} \mp$$
$$\mp \frac{1}{2}\sqrt{\left(E_d - E_b(u) + U\langle n_{p\downarrow}\rangle - t_{\bar{k}}(u)\right)^2 + 4(V(u))^2}$$

for localized electrons and standard band spectrum for itinerant ones.

In the metal insulator transition point the spectrum contain separate *d*-level with energy

$$E_1 = -\mu + E_d + U\langle n_{p\downarrow}\rangle \quad (8)$$

and band with dissipation relation

$$E_2 = -\mu + E_b(u) + t_{\bar{k}}(u). \quad (9)$$

The criterion for metal insulator transition is obtained as $E_b(u) - E_d - U\langle n_{p\downarrow}\rangle = w(u)$.

Both the bandwidth and band center position cal be changed by the external pressure application. The equilibrium value of the lattice strain can be found from a minimum condition for Gibbs function

$$G = F + PV = F + NPV_0(1 + \bar{u}),$$

as

$$\bar{u} = -\frac{1}{V_0C}\left(\frac{S}{N}\sum_{\bar{k}\sigma}\langle c^+_{\bar{k}\sigma}c_{\bar{k}\sigma}\rangle + \right.$$
$$\left. + \frac{BV_0}{2w}\frac{1}{N}\sum_{\bar{k}\sigma}t_{\bar{k}}\langle c^+_{\bar{k}\sigma}c_{\bar{k}\sigma}\rangle - \frac{PV_0}{CV_0}\right). \quad (10)$$

To calculate $\langle c^+_{\bar{k}\sigma}c_{\bar{k}\sigma}\rangle$, determined by the spectral function of band electrons, we must find the chemical potential from the condition

$$\frac{1}{N}\sum_i\left(\langle d^+_{k\sigma}d_{k\sigma}\rangle + \langle c^+_{i\sigma}c_{i\sigma}\rangle\right) = \langle n\rangle. \quad (11)$$

For the transtion to be initiated the equilibrium value of lattice strain is to be

$$\bar{u} = -(W - w_c)/(S - 0.5BV_0) < 0. \quad (12)$$

After numerical calculation with model rectangular density of states at non-zero temperature we have

$$\bar{u} = \frac{1}{CV_0}(\frac{S}{aw}\ln\left|\frac{e^{-b-a\Delta}+1}{e^{-b+a\Delta}+1}\right| - \frac{BV_0}{2w^2}\int_{-w}^{w}\frac{\varepsilon d\varepsilon}{e^{b+a\varepsilon}+1} - pV_0), \quad (13)$$

here $a = \frac{1}{\theta}\left(1+\frac{BV_0}{2w}\bar{u}\right)$ and $b = \frac{1}{\theta}\left(-\mu + E_b(\bar{u})\right)$.

Following the paper [7] we take parameter values $1/CV_0 = 0.05 eV^{-1}$; $BV_0 = -3eV$, $S$=0.4 eV; $W$=2,2 eV; $w$=2 eV; $U$=5 eV and obtain the temperature dapendence shown in figure 1 and the pressure-temperature phase diagram shown in figure 2.

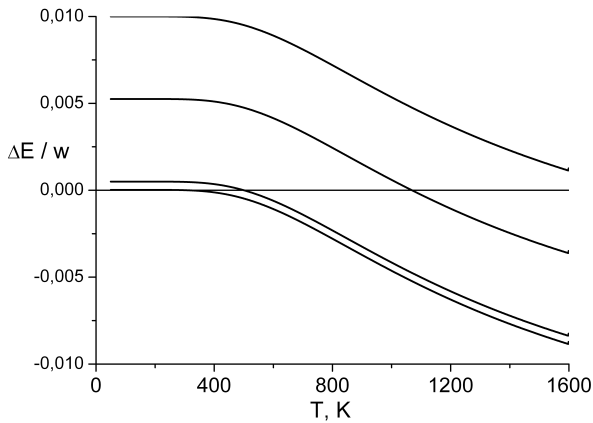

Figure 1. Energy gap width as a function of tempareture. Curves (from up to down) correspond to values $pV/w$ = 2.00, 2.05, 2.10, 2.105.

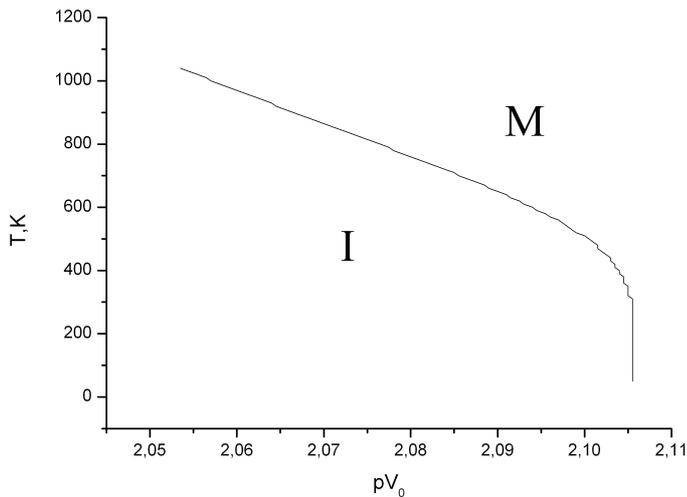

Figure 2. The pressure-temperature phase diagram of metal-insulator transition in the model.

## Conclusions

In the simplified model of narrow band compound with Anderson-Hubbard centers a stabilization of the low temperature insulator phase can be described if the effect of a lattice strain is incorporated into the model Hamiltonian. The electron localization is of Mott-Hubbard nature, in particular, the upper and lower Hubbard subbands can be observed in the energy spectrum far from the transition point. No matter how weak, the hybridization of band and localized states provide a sufficient mechanism for localization effects to dominate in a wide temperature-pressure ranges. Insulator or correlated metal phases in the phase diagram and the Coulomb repulsion-to-bandwidth or localized level energy-to-bandwith ratios, where both the bandwidth and localized levels position can be substantially changed by the external pressure or chemical substitution can in principle be estimated from the expreimental data on the basis of the considered model